\documentclass[aps,prd,twocolumn,superscriptaddress,nofootinbib]{revtex4}
\usepackage{graphicx}
\usepackage{amsmath}
\usepackage{amsfonts}
\usepackage{amssymb}

\begin{document}
\title{Probing features of the Lee-Wick quantum electrodynamics}

\author{R. Turcati}\email{turcati@cbpf.br}
\affiliation{Departamento de F\'isica e Qu\'imica, Universidade Federal do Esp\'irito Santo,
Av. Fernando Ferrari, 514, Goiabeiras, 29060-900, Vit\'oria, ES, Brazil}
\affiliation{Laborat\'{o}rio de F\'{\i}sica Experimental (LAFEX), Centro Brasileiro de Pesquisas F\'{i}sicas (CBPF), Rua Dr. Xavier Sigaud 150, Urca, 22290-180, Rio de Janeiro, RJ, Brazil}
\author{M. J. Neves}\email{mariojr@ufrrj.br}
\affiliation{Departamento de F\'isica, Universidade Federal Rural do Rio de Janeiro, BR 465-07, 23890-971, Serop\'edica, Rio de Janeiro, Brazil}

\begin{abstract}

In this paper we discuss some aspects concerning the electromagnetic sector of the abelian Lee-Wick (LW) quantum electrodynamics (QED). Using the Dirac's theory of constrained systems, the higher-order canonical quantization of the LW electromagnetism is performed. A quantum bound on the LW heavy mass is also estimated using the best known measurement of the anomalous magnetic moment of the electron. Finally it is shown that magnetic monopoles can coexist peacefully in the LW scenario. 

\keywords{Lee-Wick electrodynamics \and higher-order canonical quantization \and anomalous magnetic moment of the electron \and quantum bound on the Lee-Wick heavy mass \and magnetic monopoles}
\end{abstract}
\maketitle


\section{Introduction}
\label{intro}

In recent years, Grinstein, O'Connell and Wise extended the ideas of the LW finite QED \cite{LeeWick69,LeeWick70} to non-abelian gauge theories and gave rise to the Lee-Wick Standard Model (LWSM) \cite{Connell08}. In the LWSM, each field of the Standard Model (SM) has associated a massive LW partner, being these modes the only parameters added to the SM framework. LW theories belong to a class of higher-order gauge models that are very useful to treat ultraviolet divergences. By adding higher-order derivative kinetic terms in the lagrangian density, the modified propagator improves the behavior at high energies scale and induce the appearance of gauge massive resonances. Unlike the original LW QED, the LWSM is not finite, but is renormalizable. The reason is because the gauge covariant derivate introduce momentum dependence interactions - giving rise to degree of divergence of loops diagrams - the modified propagators have a better asymptoptic behavior in the ultraviolet range, reducing the degree of divergence in radiative corrections. By power counting arguments, these mutual effects cancel each other and give origin at most to logarithmic divergences, providing an alternative way to solve the hierarchy puzzle \cite{Connell08}. It also have been suggested that quantum gravity effects can excitate a Lee-Wick partner to every field in the SM \cite{alvarez08,wu}, which is the exactly degree of freedom required by the LWSM. In order to be consistent with the electroweak data, the LW scale must be of a few TeV \cite{alvarez08,carone08,underwood09,chivukula10}. Currently, there is a vast literature both as phenomenological issues as theoretical aspects of the LWSM \cite{fornal09,grinstein08,espinosa08,grinsconnell08,espinosa11,carone09,carone009,rodigast09,cai09,rizzo07,rizzo08,dulaney08,petrov}, which shows the growing interest in the last few years.

Since LW quantum electrodynamics is the cornerstone of the LWSM, issues concerning its properties can give a valuable insight into our understanding of the LW dynamics. A fundamental feature regarding quantum field theories (QFT) is associated to ultraviolet divergences. This shortcoming is intrinsic to interacting QFTs when excitations are point particles and interactions are local, which arise due the fact that the integrals over intermediate energies diverge at their high energy end \cite{jackiw99}. At the classical context, Podolsky \cite{podolsky42}, Podolsky and Kikuchi \cite{podolsky44,podolsky45}, Montgomery \cite{montgomery46} and Green \cite{green47,green48} developed a completely relativistic electrodynamics free from the defect of infinities self-energies and which reduces to Maxwell-Lorentz formulation for low energy phenomenon through the addition of higher-order derivative kinetic terms in Maxwell electrodynamics. This field theory has the advantage of maintaining the $U(1)$ gauge structure of the QED. In the quantum realm, J. J. Sakurai gave an outlook relative to divergence difficulties inherent to quantum theory at high energies \cite{sakurai}. In order to obtain finite values of physical observables as mass and charge of the electron in QED, the photon propagator must be modified by a cut-off parameter, as Pauli-Villars regulator for example. But the Pauli-Villars prescription is unsatisfactory since it give rise to non-Hermitian interactions and scaterring processes do not conserve probabilities, violating thereby the unitarity. Sakurai suspected that QED must be modified at short distances to overcome these difficulties at fundamental level. Soon after Sakurai's insight, Lee and Wick proposed a way to treat the QED divergences in ultraviolet range \cite{LeeWick69,LeeWick70}. The basic idea was to promote the Pauli-Villars regulator as a freedom degree in QED. The outcome is a modified photon propagator that in high frequencies limit goes as $k^{-4}$, improving the behavior of the electrodynamics at short distances. Their model is named Lee-Wick finite theory of QED. Nevertheless, a wrong sign residue appears in the poles, coming to light negative norm states in the Hilbert space and thereby breaking unitarity. Lee and Wick argued that these ghost modes should possess a heavy mass and decays in on-shell particles. It is worth to note that Lee-Wick original ideas were abandoned only after the dimensional regularization schema of gauge theories \cite{hooft72}, living a relegation era for about two decades. Nevertheless, questions concerned to the foundations of the Lee-Wick finite QED remain open -  until today, for example, do not exist a demonstration concerning the unitarity at arbitrary loops in perturbative formalism; however no exceptions were found and satisfactory answers are not yet settled.

Following up previous works \cite{accioly10,accioly11}, the focus of this paper is on the electromagnetic sector of the LW quantum electrodynamics and it is organized as follows. In Section II we review some properties of the abelian Lee-Wick electrodynamics. In Section III the higher-order canonical quantization is considered in detail. In Section IV we estimate a quantum bound in the LW heavy mass using the experimental value of the anomalous magnetic moment of the electron. In Section V we discuss about the possibility of coexistence of magnetic monopoles and the duality simmetry in the LW model. Our conclusions are presented in Section VI.

In our conventions $\hbar=c=1$ and the signature of the metric is ($+1,-1,-1,-1$).


\section{Overview of the abelian Lee-Wick model}
\label{sec:1}
The Abelian LW model is defined by the following gauge-invariant Lagrangian
\begin{equation}\label{lwes}
\mathcal{L}=-\frac{1}{4}F_{\mu\nu}F^{\mu\nu}-\frac{1}{4M^2}F_{\mu\nu}\Box F^{\mu\nu},
\end{equation}
where $F_{\mu \nu} (= \partial_\mu A_\nu - \partial_\nu A_\mu)$ is the field strength.

Let us then show that the above Lagrangian describes two independent (on-shell) spin-1 fields:  massless one and  massive one, with positive and negative norm, respectively. To do that it is appropriate to provide another formulation where an auxiliary field is introduced and the higher-derivative term is absent. The field theory with real vectorial fields $A_\mu$ and $Z_\mu$ with Lagrangian
\begin{eqnarray}\label{lwsv}
{\cal{L}} &=& \frac{1}{2}A_\mu \Box Z^\mu + \frac{1}{2}\partial_\mu A^\mu \partial_\nu Z^\nu - \frac{M^2}{8}A_\mu A^\mu \nonumber \\ &&+ \frac{M^2}{4}A_\mu Z^\mu - \frac{M^2}{8}Z_\mu Z^\mu,
\end{eqnarray}
is equivalent to the field theory with the Lagrangian in Eq. (\ref{lwes}). In fact, varying $Z_\mu$ gives 
\begin{eqnarray}  
Z_\mu = A_\mu + \frac{2}{M^2}\Box A_\mu - \frac{2}{M^2}\partial_\mu \partial_\nu A^\nu,
\end{eqnarray}
and the coupled second-order equations from (\ref{lwsv}) are fully equivalent to the fourth-order equations from (\ref{lwes}). The system (\ref{lwsv}) now separates cleanly into the Lagrangians for two fields, when we make the change of variables
\begin{eqnarray}
A_\mu = B_\mu + C_\mu, \\ Z_\mu= B_\mu - C_\mu.
\end{eqnarray}

In terms of $B_\mu$, $C_\mu$, $B_{\mu \nu} \equiv \partial_\mu B_\nu - \partial_\nu B_\mu$ and $C_{\mu \nu } \equiv \partial_\mu C_\nu - \partial_\nu C_\mu$, the Lagrangian now becomes
\begin{eqnarray}\label{lwsf}
{\cal{L}} = -\frac{1}{4}B_{\mu \nu}B^{\mu \nu} + \frac{1}{4}C_{\mu \nu}C^{\mu \nu} - \frac{M^2}{2}C_\mu C^\mu,
\end{eqnarray}
which is nothing but the difference of the Maxwell Lagrangian for $B_\mu$ and the Proca Lagrangian for $C_\mu$. 

The particle content of the theory can also be obtained directly from Eq. (\ref{lwes}). To accomplish this goal we compute the residues at the simple poles of the saturated propagator (contraction of  the propagator with conserved currents). Adding to (\ref{lwes}) the gauge-fixing term ${\cal{L}}_\lambda = -\frac{1}{2\lambda}(\partial_\mu A^\mu)^2$, where as usual $\lambda$ plays the role of the gauge-fixing parameter, and noting that due to the  structure of the theory and the choice of a linear gauge-fixing functional, no Faddeev-Popov ghosts are required in this case, we promptly get the propagator in momentum space, namely,
\begin{eqnarray}\label{plwt}
D_{\mu \nu}(k) &=& \frac{M^2}{k^2(k^2- M^2)}\left\{\eta_{\mu \nu} - \frac{k_\mu k_\nu }{k^2} \Bigg[1\right.\nonumber\\
&&\left.+\lambda \left(\frac{k^2}{M^2} - 1  \right) \Bigg]\right\}.
\end{eqnarray}

Contracting (\ref{plwt}) with conserved currents $J^\mu(k)$, yields 
\begin{eqnarray}
{\cal{M}} \equiv J^\mu D_{\mu \nu} J^\nu, \nonumber \; \;= -\frac{J^2}{k^2} + \frac{J^2}{k^2 - M^2},
\end{eqnarray}
which allows us to conclude, taking into account that $J^2 < 0$ \cite{accioly03,accioly04,accioly05}, that the signs of the residues of ${\cal{M}}$ at the poles $k^2=0$ and $k^2 = M^2$ are, respectively, 
$$Res {\cal{M}}(k^2=0) >0, \;\;\; Res{\cal{M}}(k^2=M^2)<0,$$ 
which confirms our previous result.

It is worth noticing that the wrong sign of the residue of the heavy particle indicates an instability of the theory at the classical level. From the quantum point of view it means that the theory is nonunitary. Luckily, these difficulties can be circumvented. Indeed, the classical instability can be removed by imposing a future boundary condition  in order to prevent exponential  growth of certain modes. However, this procedure leads to causality violation in the theory \cite{coleman}; fortunately, this acausality is suppressed below the scales associated with the LW particles. On the other hand, Lee and Wick argued that despite the presence of the aforementioned degrees of freedom associated with a non-positive definite norm on the Hilbert space,  the theory could nonetheless be unitary as long as the new LW particles obtain decay widths. There is no general proof of unitary at arbitrary loop order for the LW electrodynamics; nevertheless, there is no known example of unitarity violation. Accordingly, the LW electrodynamics is finite. Therefore, we need not be afraid of the massive spin-1 ghost.   

In summary, we may say that  the LW work consists essentially in  the introduction of  Pauli-Villars, wrong-sign propagator, fields as physical degrees of freedom which leads to amplitudes that are better behaved in the ultraviolet and render the logarithmically divergent QED finite.  

We remark that for the sake of convenience we shall work in the representation of the gauge field $A_\mu$ as given in Eq. (\ref{lwes}), with the propagator as in Eq. (\ref{plwt}).  


\section{Lee-Wick Canonical Quantization}
\label{sec:2}

Higher-order canonical quantization have been performed for some authors in the past \cite{pimentel88,natividade91,dutra94,bufalo11}. Nevertheless, some questionable results are presented in these works. We initially will accomplish the canonical quantization of the Lee-Wick electromagnetism and at the end of this section discuss these controversial outcomes. Our starting point for the higher-order canonical quantization is the following LW lagrangian density
\begin{equation}\label{LM}
\mathcal{L}=-\frac{1}{4}F_{\mu\nu}^{2}+\frac{1}{2M^{2}}\partial_{\mu}F_{\alpha\beta}\partial^{\mu}F^{\alpha\beta} \; .
\end{equation}

We shall analyze (\ref{LM}) instead of (\ref{lwes}) since these lagrangians are the same up to a total derivate. This choice is due to the fact that the lagrangian (\ref{lwes}) have third order derivative fields, which introduce additional complications in the canonical quantization. The pairs of canonically conjugate variables related to lagrangian (\ref{LM}) are $(A^{\alpha},\pi_{\alpha})$ and $(\bar{A}^{\alpha},\eta_{\alpha})$ respectively, where $\bar{A}^{\alpha}\equiv\dot{A}^{\alpha}$ is an independent variable. Since gauge invariance holds in the LW model, the second-order lagrangian (\ref{LM}) is degenerate, i.e., the Hessian matrix is singular.

The set of generalized canonical momenta are
\begin{eqnarray}
\label{c1}
&&\pi^{\nu}=-F^{0\nu}+\frac{1}{M^2}\left(\Box{F^{0\nu}}+\partial_{0}\partial_{i}F^{i\nu}\right) \; , \\
\label{c2}
&&\eta^{\nu}=\frac{1}{M^2}\partial_{0}F^{0\nu}.
\end{eqnarray}

Primary constraints are obtained from the definition of the canonical momenta, without making use of the equations of motion. According to relations (\ref{c1}) and (\ref{c2}), the LW model has the following primary constraints:
\begin{eqnarray}
\label{pc1}
&&\eta_0\approx 0 \; , \\
\label{pc2}
&&\pi_0+\partial_{i}\eta_{i}\approx0 \; ,
\end{eqnarray}
where ``$\approx$" means weak equations according to Dirac's method \cite{dirac50,review}. It is necessary evaluate the dynamics of the constraints. As usual in hamiltonian formalism, it is required compute the canonical hamiltonian, which is given by
\begin{eqnarray}
\nonumber H_{C}&=&\int d^3\mathbf{x}\left[\pi^{\alpha}\bar{A}_{\alpha}+\eta^{\alpha}\dot{\bar{A}}-\mathcal{L}\right]\\
\nonumber &=&\int d^3\mathbf{x} \left[\pi_{0}\bar{A}_0-\pi_{i}\bar{A}_{i}+\bar{A}_0(\partial_{i}\eta_{i})
-\frac{1}{2}M^{2}\eta^2_{i}\right. \nonumber\\
&&+(\partial_{j}\eta_{i})F_{ji}-\frac{1}{2}F_{0i}^{2}+\frac{1}{4}F_{ij}^2-\frac{1}{2M^{2}}\partial_{i}F_{i0}\partial_{j}F_{j0}\nonumber \\
&&\left.-\frac{1}{4M^{2}}\partial_{0}F_{ij}\partial_{0}F_{ij}+\frac{1}{4M^{2}}\partial_{k}F_{ij}\partial_{k}F_{ij}\right].\;\;\;
\end{eqnarray}

The primary hamiltonian $H_{1}$ is defined as
\begin{equation}
H_{1}=H_{C}+\int d^3\mathbf{x}\left[\lambda_{1}\eta_{0}+\lambda_{2}(\pi_0+\partial_{i}\eta_{i})\right].
\end{equation}

So, applying the Poisson brackets we obtain
\begin{eqnarray}\label{c3}
&&\dot{\eta}_{0}=\{\eta_0,H_{1}\}\approx0 \; , \\ \label{c4}
&&(\dot{\pi}_0+\partial_{i}\dot{\eta}_{i})=\{\pi_0+\partial_{i}\eta_{i},H_{1}\}=\partial_{i}\pi^{i}\approx0 \; ,
\end{eqnarray}
which reveal us the appearance of a non-primary constraint. The secondary constraint arise from the condition that the primary constraints should be preserved in time. It is necessary identify all the constraints of the model. Then, computing the Poisson brackets again, but now with the secondary hamiltonian
\begin{equation}
H_{2}=H_{C}+\int d^3\mathbf{x}\left[\lambda_{1}\eta_0+\lambda_2(\pi_0+\partial_{i}\eta_{i})+\lambda_3\partial_{i}\pi_{i}\right],
\end{equation}
we get $\partial_{i}\dot{\pi}_{i}=\{\partial_{i}\pi_{i},H_{2}\}\approx0$. No more constraints appear in our formalism, which implies that the consistency condition is identically fulfilled. 

Analyzing the Hamilton field equations of motion to $A_{0}$ and $A_{i}$ provide us
\begin{eqnarray}
\dot{A}_{0}&=&\{A_0,H\}=\bar{A}_0+\lambda_2,\\
\dot{A}_{i}&=&\{A_{i},H\}=\bar{A}_{i}+\partial_{i}\lambda_3,
\end{eqnarray}
which allow us choose the following lagrange multipliers:
\begin{equation}
\lambda_2=0,\quad \lambda_3=0.
\end{equation}

On the other hand, the equations of motion concerning to $\bar{A}_{i}$ and $\bar{A}_0$ yields
\begin{eqnarray}
\label{ec1}
\dot{\bar{A}}_{i}&=&\{\bar{A}_{i},H\}=M^{2}\eta_{i}+\partial_{i}\bar{A}_0,\\
\label{ec2}
\dot{\bar{A}}_0&=&\{\bar{A}_0,H\}=\lambda_1.
\end{eqnarray}

Equation (\ref{ec1}) is nothing but $\eta^{i}=\frac{1}{M^2}\partial_{0}F^{0i}$. The function related to the $\lambda_1$ is found through equation (\ref{ec2}). The equations related to $\eta_{j}$, $\pi_{0}$ and $\pi_{j}$ do not introduce further information. The extended hamiltonian can be expressed as $H_{E}=H_{C}+\int d^3\mathbf{x}\dot{\bar{A}}_0\eta_0$, which give us
\begin{eqnarray}
H_{E}&=&\int d^3\mathbf{x} \left[\pi_{0}\bar{A}_0-\pi_{i}\bar{A}_{i}+\bar{A}_0(\partial_{i}\eta_{i})
-\frac{1}{2}M^{2}\eta^2_{i}\right. \nonumber\\
&&+(\partial_{j}\eta_{i})F_{ji}-\frac{1}{2}F_{0i}^{2}+\frac{1}{4}F_{ij}^2-\frac{1}{2M^{2}}\partial_{i}F_{i0}\partial_{j}F_{j0}\nonumber \\
&&\left.-\frac{1}{4M^{2}}\partial_{0}F_{ij}\partial_{0}F_{ij}+\frac{1}{4M^{2}}\partial_{k}F_{ij}\partial_{k}F_{ij}+\dot{\bar{A}}_0\eta_0\right].\nonumber \\
\end{eqnarray}

The constraints obtained are all of the first class, i.e., the Poisson bracket with all the other constraints vanishes identically, which is a direct consequence of $U(1)$ gauge invariance. The canonical quantization requires that we impose a gauge choice and remove the non-physical variables. The gauge condition necessary to change the set of the first class into the second class constraints is obtained by analysing the LW field equations in terms of potential $A^{\mu}$, which can be expressed as
\begin{equation}\label{lwc}
\left(1+\frac{\Box}{M^2}\right)\Box A^{\nu}-\partial^{\nu}\left(1+\frac{\Box}{M^2}\right)\partial_{\mu}A^{\mu}=0 \; .
\end{equation}

The equation (\ref{lwc}) suggests that a possible gauge choice is
\begin{equation}
\left(1+\frac{\Box}{M^2}\right)\partial_{\mu}A^{\mu}=C \; ,
\end{equation}
where $C$ is a arbitrary constant which can be choosen equal to zero, which provide us the following gauge conditions:
\begin{equation}
\bar{A}_0=0 \; , \quad \left(1+\frac{\Box}{M^2}\right)\nabla\cdot\mathbf{A}=0\; , \quad A_0=0 \; .
\end{equation}

The LW field equation (\ref{lwc}) is compatible with this particular gauge choice \footnote{We could also choose the gauge conditions $\bar{A}_0=0$, $\nabla\cdot\mathbf{A}=0$, $A_0=0$.}. The gauge constraints also satisfy the consistency condition. All the constraints obtained are now of the second class and are given by
\begin{eqnarray}
\Omega_1\!\!\!&=&\!\!\!\eta_0\approx0,\quad \Omega_2=\pi_0+\partial_{i}\eta_{i}\approx0,\quad \Omega_3=\partial_{i}\pi_{i}\approx0 \, ,
\nonumber \\
\Omega_4\!\!\!&=&\!\!\!\bar{A}_0\approx0,\quad \Omega_5=\left(1+\frac{\Box}{M^2}\right)\partial_{i}A_{i}\approx0,\quad \Omega_6=A_0\approx0 \; . \nonumber
\end{eqnarray}

The set of constraints found enable us to determine the constrained matrix. The only elements of the $C_{\alpha\beta}\equiv\{\Omega_{\alpha},\Omega_{\beta}\}$ nonzero are
\begin{eqnarray*}
\{\Omega_{1},\Omega_{4}\}&=&-\{\Omega_{4},\Omega_{1}\}=\delta^{3}(\mathbf{x}-\mathbf{x'}), \\
\{\Omega_{2},\Omega_{6}\}&=&-\{\Omega_{6},\Omega_{2}\}=\left(1-\frac{\nabla^{2}}{M^{2}}\right)\nabla^{2}\delta^{3}(\mathbf{x}-\mathbf{x'}), \\
\{\Omega_{3},\Omega_{5}\}&=&-\{\Omega_{5},\Omega_{3}\}=\delta^{3}(\mathbf{x}-\mathbf{x'}). 
\end{eqnarray*}

Since the constraints are all of second class, the $C$-matrix is invertible and the Dirac brackets are completely charactarized by
\begin{eqnarray}
&&\{A(x),B(x')\}_{D}=\{A(x),B(x')\}\\ \nonumber
&&-\int{d}^{3}\mathbf{y}d^{3}\mathbf{z}\{A(x),\Omega_{a}(y)\}C_{ab}^{-1}(y,z)\{A(x),B(x')\},
\end{eqnarray}
Computing the inverse $C$-matrix and taking into account that the green function $G(\mathbf{x}-\mathbf{x'})$ satisfies the equation $\left(1-\frac{\nabla^{2}}{M^{2}}\right)\nabla^{2}G(\mathbf{x}-\mathbf{x'})=-\delta^{3}(\mathbf{x}-\mathbf{x'})$ and is given by
\begin{equation}
G(\mathbf{x}-\mathbf{x'})=\frac{1}{4\pi}\frac{1}{|\mathbf{x}-\mathbf{x'}|}\left[1-e^{-M|\mathbf{x}-\mathbf{x'}|}\right],
\end{equation}
the Dirac brackets then becomes
\begin{eqnarray*}
\{A^{\mu}(\mathbf{x},t),\pi^{\nu}(\mathbf{x'},t)\}_{D}=(\eta^{\mu\nu}-\eta^{\mu0}\eta^{\nu0})\delta^3(\mathbf{x-x'})\\
+\eta^{\mu i}\eta^{\nu j}\left(1-\frac{\nabla^2}{M^{2}}\right)\frac{\partial}{\partial x^{i}}\frac{\partial}{\partial x'^{j}}G(\mathbf{x-x'}),\\
\\
\{\bar{A}^{\mu}(\mathbf{x},t),\eta^{\nu}(\mathbf{x'},t)\}_{D}=(\eta^{\mu\nu}-\eta^{\mu0}\eta^{\nu0})\delta^3(\mathbf{x-x'}).
\end{eqnarray*}

According to Dirac's method, we must write these equations as strong equalities. Using the canonical quantization prescription ($\{A,B\}_{D}\rightarrow-i[A,B]$), the LW commutators are given by
\begin{eqnarray}
\label{db}
\left[A_{i}(\mathbf{x},t),\pi_{j}(\mathbf{x},t)\right]&=&i\delta_{ij}\,\delta^3(\mathbf{x-x'})\nonumber \\
&&+i\left(1-\frac{\nabla^2}{M^2}\right)\partial_{i}\partial'_{j}G(\mathbf{x-x'}),\nonumber\\ \\  \; 
\label{db1}
\left[\bar{A}_{i}(\mathbf{x},t),\eta_{j}(\mathbf{x'},t)\right]&=&i\delta_{ij}\,\delta^3(\mathbf{x-x'}) \;.
\end{eqnarray}

It is worth noting that in the absence of higher-order derivative terms, the commutator (\ref{db}) reproduces the Maxwell commutation relation. To end up, is important to note that the Poincar\'e algebra is also satisfied.

The history of the quantization of higher-order electromagnetic theories began a long time ago. The first who tried to perform the higher-order quantization in the electromagnetism was B. Podolsky and Kikuchi in 1944 \cite{podolsky44} and some years later B. Podolsky and P. Schwed \cite{podolsky48} using the Glupta-Bleuer method; the results obtained however are dubious due to the fact that in that epoch there was not an easy method to deal with quantization methods of gauge theories, a procedure not complete clear up to now. In 1950s, G. R. Pitman \cite{pitman55} and R. E. Martin \cite{martin60} probed some aspects of Podolsky electrodynamics in their Ph.D. thesis. Again, the results found are not correct due the inability of treating the electron self-energy. 

Higher order canonical quantization by Dirac formalism began in early $1960$'s. Since then, some authors tried to achieve the quantization of higher-order theories \cite{barcelos89,braga89,barcelos91,muslin00} via Dirac's method; on the other hand, canonical quantization of higher-order electromagnetism starting from the determination of primary and secondary constraints have been performed in references \cite{pimentel88,natividade91,dutra94,bufalo11}. Nevertheless, there are some misconceptions in previous works. In \cite{pimentel88} and \cite{natividade91} the density lagrangian is incorrect. On the other hand, in \cite{pimentel88} the number of primary constraints is 2, while in \cite{natividade91} the same theory discussed in \cite{pimentel88} introduces only one primary constraint. As previously stated, primary constraints follow solely from the definition of the canonical momenta while non-primary constraints arise directly from the condition that the primary constraints hold in time \cite{dirac50,review}. In \cite{pimentel88} and \cite{bufalo11} the authors argue that cannot be used the \emph{gauge} $\nabla\cdot\mathbf{A}=0$, but only $\left(1+\frac{\Box}{M^2}\right)\nabla\cdot\mathbf{A}=0$. In \cite{natividade91} in turn, it is claim that both gauges are feasible. Upon gauge fixation, the set of the first class constraints and gauge conditions turn into a second class constraints \cite{dirac50,review}. Any gauge condition can be used provided it is consistent with field equations. We hope with this brief discussion close the misunderstanding concerning to the higher-order canonical quantization. We still emphasize that the lagrangian (\ref{LM}) used for us is not the same as the previous authors used. Nevertheless, the physical content is completely equivalent, as it should be.


\section{Anomalous magnetic moment of the electron and the LW heavy mass}
\label{sec:3}

\begin{figure}[!hbp]
\begin{center}
\includegraphics[scale=0.58]{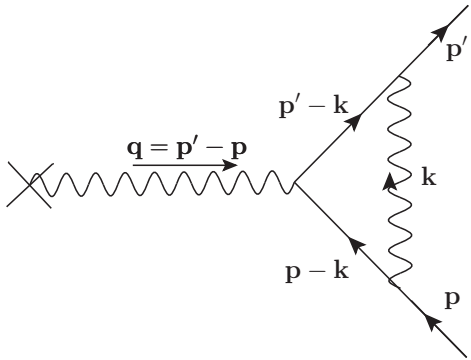}
\end{center}
\caption{\small Vertex correction for electron  scattering by  an external  field.\label{fig1}}
\end{figure}

Taking into account that QED predicts the anomalous magnetic moment of the electron correctly to ten decimal places, a quantum bound on the mass $M$ of the LW heavy particle can be found by computing the anomalous electron magnetic moment  in the context of the LW electrodynamics and   comparing afterward the result obtained with that of QED. To accomplish this goal, we recall that the anomalous magnetic moment  stems from the vertex correction for the scattering of the electron by an external field, as it is shown in Fig. (\ref{fig1}).

For an electron scattered by an external static magnetic field and in limit ${\bf q} \rightarrow {\bf 0}$, the gyromagnetic ratio is given by \cite{frampton}
\begin{equation}
\label{eamm}
g=2[1+F_{2}(0)]. 
\end{equation}

The form factor of the electron, $F_2(0)$, corresponds to a shift in the $g-$factor, usually quoted in the form $F_2(0)\equiv \frac{g-2}{2}$, and yields the anomalous magnetic moment of the electron.  By employing (\ref{eamm}) in the calculation of the diagram in Fig. (\ref{fig1}), it can be shown that 
\begin{eqnarray}
F_{2}(0)&=&\frac{\alpha}{\pi}\int_0^\infty d\alpha_1 d\alpha_2 d \alpha_3 \delta (1 - \Sigma \alpha_i) \left[\frac{\alpha_1}{\alpha _2 + \alpha_3}\right. \nonumber \\ 
&&\left.-\frac{\alpha_1^2 (\alpha_2 + \alpha_3)}{(\alpha_2 + \alpha_3)^2 + \frac{\alpha_1}{\varepsilon}}\right],
\end{eqnarray}
where $\varepsilon \equiv \frac{m^2}{M^2}$, $m$ being the electron mass. We call attention to the fact that the term $-\frac{M^2k^\mu k^\nu}{k^4(k^2-M^2)}\left[1+\lambda\left(\frac{k^2}{M^2}-1\right)\right]$ that appears in Eq. (\ref{plwt}) makes  no  contribution to the the form factor $F_{2}(0)$ because the propagator always occurs coupled to conserved currents. 

Integrating the above expression  first with respect to $\alpha_3$ and subsequently with respect to $\alpha_2$, gives 
\begin{eqnarray}
F_2 (0) &=& \frac{\alpha}{\pi}\int_0^1 d\alpha_1 \int_0^{1- \alpha_1} d \alpha_2\left[\frac{\alpha_1}{1- \alpha_1}-  \frac{\alpha_1 (1 - \alpha_1)}{(1- \alpha_1)^2 + \frac{\alpha_1}{\varepsilon}} \right] \nonumber \\ &=& \frac{\alpha}{\pi}\int_0^1 d \alpha_1 \frac{\alpha_1^2}{\alpha_1 + \varepsilon (1 -\alpha_1)^2}. 
\end{eqnarray}

Computing $F_{2}(0)$, we arrive at the conclusion that
\begin{eqnarray}\label{lwsoc}
F_{2}(0)&\approx&\frac{\alpha}{2\pi}\left[1-\frac{2}{3}\left(\frac{m}{M}\right)^2 - 2\left( \frac{25}{12} +  \ln \left(\frac{m}{M}\right) \right)\left(\frac{m}{M}\right)^4 \right. \nonumber \\
&&\left.+ {\cal O}\left(\left(\frac{m}{M}\right)^6\right)\right].
\end{eqnarray}

The first term of the above equation is equal to that calculated by Schwinger in 1948 \cite{schwinger48}. Since then  $F_2(0)$ has been calculated to order $\alpha^{10}$ for QED. The second term of Eq. (\ref{lwsoc}) is the most important correction related to the parameter $M$ of the LW electrodynamics.

Recent calculations  concerning $F_2(0)$ in the framework of QED give for the  electron \cite{aoyama12}
\begin{equation}
F_2(0) = 1 \;159 \;652\; 181.78\,(0.06)(0.04)(0.02)\times 10^{-12} \; ,\nonumber
\end{equation}
where the uncertainty comes mostly from that of the best non-QED value of the  fine structure constant $\alpha$. The current experimental value for the anomalous magnetic moment is, in turn, \cite{hanneke08}
\begin{equation}
F_2(0) = 1 \;159 \;652\; 180.73\,(0.28)\times 10^{-12} \, . \nonumber
\end{equation}

Comparison of the theoretical value predicted by QED  with the experimental one shows that these results agree in $1$  part in  $10^{12}$. As a consequence, 
\begin{equation}\label{qblw}
\frac{2}{3}\left(\frac{m}{M}\right)^2<10^{-12}. 
\end{equation}

Consequently, a lower limit on the heavy particle Lee and Wick hypothesized the existence is $M\approx$ 409 GeV.

%


\section{Duality Symmetry and Monopoles in the Lee-Wick Electrodynamics}
\label{sec:4}
An important question concerning the LW finite QED is whether or not the LW heavy particle and magnetic charge can live in peace in its context. To answer this question we introduce a magnetic current $J_{m}^{\mu}=(\rho_{m},\mathbf{J}_{m})$ in the LW dual field equations. It is fairly straightforward to show that the resulting system of modified higher-order field equations, namely,
\begin{eqnarray}\label{lwf1}
\left(1+\frac{\Box}{M^{2}}\right)\partial_{\mu}F^{\mu\nu}&=&J_{e}^{\nu} \\
\label{lwf}
\partial_{\mu}\tilde{F}^{\mu\nu}&=&J_{m}^{\nu}
\end{eqnarray}
where $\tilde{F}^{\mu\nu}=\frac{1}{2}\epsilon^{\mu\nu\alpha\beta}F_{\alpha\beta}(\epsilon^{0123}=+1)$, describes the existence of a magnetic charge. In fact, assuming the absence of electric fields, charges, and currents as well as the absence of magnetic current, we are left essentially with two equations for the magnetic field which have the familiar Dirac monopole solution $\mathbf{B}=\frac{q_{m}}{4\pi{r}^{2}}\mathbf{\hat{r}}$, where $q_{m}$ is the magnetic charge. Using the usual methods, the Dirac quantization condition $\frac{q_{e}q_{m}}{4\pi}=\frac{n}{2}$, where $q_{e}$ is the electric charge, and $n$ is an integer, can be promptly recovered. We have thus succeeded in finding a consistent system of Maxwell + vectorial boson mass + magnetic charge equations. We remark that the Dirac monopole and the massive vectorial boson cannot coexist in the context of Proca massive electrodynamics \cite{ignatiev96} because the latter, unlike the LW QED, is not gauge invariant. The very existence of the Dirac monopole is undoubtedly linked to the existence of the gauge invariance of the corresponding theory. Interestingly enough the system formed by Eqs. (\ref{lwf1}) and (\ref{lwf}) is not symmetric under the duality transformation $F^{\mu\nu}\rightarrow\tilde{F}^{\mu\nu}$, $\tilde{F}^{\mu\nu}\rightarrow-{F}^{\mu\nu}$, augmented by $J_{e}^{\mu}\rightarrow {J}_{m}^{\mu}$, $J_{m}^{\mu}\rightarrow-{J}_{e}^{\mu}$. This fact raises an interesting question: Would it be possible to accomodate simultaneously magnetic charge and duality transformations in the framework of a higher-order electromagnetic model? A good attempt in this direction might be, for instance, the model defined by the field equations
\begin{eqnarray}
\left(1+\frac{\Box}{M^{2}}\right)\partial_{\mu}F^{\mu\nu}&=&J_{e}^{\nu} \\
\label{lwg}
\left(1+\frac{\Box}{M^{2}}\right)\partial_{\mu}\tilde{F}^{\mu\nu}&=&J_{m}^{\nu}
\end{eqnarray}
since it is symmetric under duality transformations. It is worth noticing that $\left(1+\frac{\Box}{M^{2}}\right)\partial_{\mu}\tilde{F}^{\mu\nu}=J_{m}^{\nu}$ is identically zero in the absence of the magnetic current. Let us see then whether this model admits a monopole-like solution. For a magnetostatic charge of strength $q_{m}$ fixed at the origin the solution of the preceding equations is
\begin{eqnarray}\label{Bmonopolo}
\mathbf{B}=\frac{q_{m}}{4\pi}\left[\frac{1-\left(1+Mr\right)e^{-Mr}}{r^{2}}\right]\mathbf{\hat{r}} \; ,
\end{eqnarray}
which for large distances reduces to the Dirac result, as it should be. Our point, nonetheless, is to ascertain whether or not this solution describes a magnetic monopole at short distances. To see this we calculate the flux of the radial magnetic field through a spherical surface $\partial{\cal R}$ of radius $r$ with the static monopole of strength $q_{m}$ at its center. Performing the computation we promptly find $\oint_{\partial{\cal R}}\mathbf{B} \cdot {\bf \hat{n}}\,{dA}= q_{m}\left[1-\left(1+Mr\right)e^{-Mr}\right]$
which implies that for $Mr<<1$,$\oint_{\partial{\cal R}}\mathbf{B}\cdot {\bf \hat{n}}\,{dA}\approx0$.
Now, taking into account that if $\mathbf{B}=\mathbf{\nabla}\times\mathbf{A}$, $\oint_{\partial{\cal R}}\mathbf{B}\cdot {\bf \hat{n}}\,{dA}$ vanishes identically, we come to conclusion that $\mathbf{A}$ can exist everywhere in the region under consideration, which show us that in the short range limit the Dirac quantization condition cannot be recovered. To see if this actually occurs, we take into account that for $Mr<<1$, $\mathbf{B}\approx\frac{q_{m}M^{2}}{8\pi}\hat{\mathbf{r}}$, implying that the magnetic field is constant at short distances instead of falls down with $\frac{1}{r^{2}}$. This bizarre behavior of the magnetic field certainly precludes us from recovering the Dirac quantization condition. One heuristic way of seeing that is to consider the motion of a particle of mass m and charge $q_{e}$ in the field of the magnetic monopole.  
From the equation of motion of the eletric charge, $m\mathbf{\ddot{r}}=q_{e}\mathbf{\dot{r}}\times\mathbf{B}$, we get the ratio of change of its angular momentum
\begin{equation}
\frac{d}{dt}\left(\mathbf{r}\times{m}\mathbf{\dot{r}}\right)=\frac{q_{e}q_{m}M^{2}r^{2}}{8\pi}\frac{d\mathbf{\hat{r}}}{dt} 
\end{equation}
a result that prevents us from defining a conserved total angular momentum as in the case of the Dirac monopole. Now, if the distances are neither too large nor much small, the potential vector cannot exist everywhere in the domain bounded by ${\partial{\cal R}}$ because $F^{\mu\nu}$ satisfies Eq. (\ref{lwg}) rather than Eq. (\ref{lwf}). Unlucky we cannot overcame this difficult by introducing the concept of a string as Dirac did since in this case $\mathbf{\nabla}\cdot\mathbf{B}=\frac{q_{m}M^{2}}{4\pi}\frac{e^{-Mr}}{r}$ does not vanishes anywhere in the aforementioned domain. The preceding analysis leads us to conjecture that Dirac-like monopoles and duality transformations cannot be accommodated in the context of one and same higher-order electromagnetic model.


\section{Final Remarks}

In this paper we have studied many issues of the electromagnetic sector of the LW model. First, we saw that questions related to higher-order canonical quantization can be very tricky, as for instance, the definition of primary and non-primary constraints as well the gauge conditions. An extensive argumentation why it is the correct way to proceed in higher-order canonical quantization was accomplished at the end of the section III.

About magnetic monopoles, a possible road of investigation is concerning to Lorentz violation theories. One interesting theory to evaluate in this scenario is the Myers-Pospelov (MP) model. MP is an effective higher-order gauge invariant theory that violates the Lorentz symmetry in the electromagnetic sector. It would be interesting analyze the eventual existence of magnetic monopoles and the presence of dual symmetry in a model that violates the lorentz symmetry together with higher-order derivative terms. Other field of great interest would be the search for monopoles in the non-abelian generalization of Lee-Wick QED. In the context of LWSM the search for monopoles can be achieved by finding topologically non-trivial finite-energy solutions.

To conclude, the bound we have found on the LW heavy mass was obtained using the most accurate experimental data currently available as input for the anomalous magnetic moment of the electron. As far as the truly (loop) quantum effects are concerned, a quick glance at equation (\ref{qblw}) clearly shows that a better agreement between theory and experiment concerning the anomalous magnetic moment of the electron would lead to an improvement of the quantum bound. Consequently, there is a great probability of setting a better quantum bound on the LW heavy mass in the foreseeable future.


\begin{acknowledgements}
Rodrigo Turcati is very grateful to CNPq (Brazilian agency) for financial support. The authors are very grateful to the referees for their helpful suggestions and comments. 
\end{acknowledgements}



\end{document}